**Title**: Temporal filling-in reduces attentional fluctuations in sustained visual attention

Running head: Temporal filling-in reduces attentional fluctuations

**Author**: Yingyu Huang [1], Liying Zhan [1,2], Xiang Wu [1*]

[1]Department of Psychology, Sun Yat-Sen University, Guangzhou, China

[2]School of Education, Zhaoqing University, Zhaoqing, China

* Correspondence:

Xiang Wu

Department of Psychology, Sun Yat-Sen University, 132 Waihuan East Road, Higher Education Mega Center, Guangzhou, Guangdong, China, 510006

E: wuxiang3@mail.sysu.edu.cn

## Abstract

Our capacity to maintain focus on task-relevant goals over time is constrained because attentional states wax and wane moment to moment. One theoretical account posits that temporal filling-in – filling temporal blank intervals between target stimuli – reduces attentional fluctuations to enhance sustained attention. This proposal, however, lacks direct empirical validation given the difficulty of tracking attentional fluctuations via behavioral measurements. To resolve this gap, we combined a visual timing task that assessed the behavioral benefit of temporal filling-in with electroencephalographic recordings of visual steady-state evoked potentials (SSEPs), which tracked neural responses associated with periodic attentional fluctuations entrained to the rhythmic visual sequence. Behavioral improvements in sustained attention via temporal filling-in were associated with reduced attentional fluctuations, as indexed by attenuated visual SSEP amplitudes. The findings highlight a fundamental neural mechanism whereby temporal filling-in reduces attentional fluctuations to stabilize sustained visual attention over extended periods.

**Significance**:

Sustained visual attention is disrupted by moment-to-moment attentional fluctuations. One theoretical account posits that temporal filling-in – filling blank intervals between target stimuli – mitigates such fluctuations, yet this proposal lacks direct empirical evidence, since behavioral measurements fail to capture real-time attentional shifts. We combined a rhythmic visual timing task with visual steady-state evoked potential (SSEP) recordings to track neural responses associated with periodic attentional fluctuations. Temporal filling-in improved behavioral timing performance and attenuated visual SSEP amplitudes. Our findings uncover a fundamental neural mechanism through which temporal filling-in stabilizes sustained visual attention by suppressing attentional fluctuations.

## Introduction

When sustaining attention to task-relevant goals over a period of time, attentional states fluctuate continuously from moment to moment (Sarter et al., 2001; Langner and Eickhoff, 2013; Fortenbaugh et al., 2017; Esterman and Rothlein, 2019). Sustained attention was traditionally investigated via detection or discrimination of infrequent targets among frequent non-targets, such as vigilance tasks and continuous performance tasks (CPTs). Recent work has expanded to other paradigms addressing broader questions of sustained attention (Fortenbaugh et al., 2017). One prominent line of rhythmic temporal processing research involves periodic attentional fluctuations (Large and Jones, 1999; Repp, 2005), including tasks requiring detection of timing deviations within a temporally regular sequence (i.e., a metronome) (O'Connell et al., 2009) or synchronized finger tapping to a metronome beat (or pulse) (Hove et al., 2017; Laflamme et al., 2018).

Sustaining focused attention is inherently challenging because target stimuli are not always continuously presented over time, and sustained attention paradigms typically present a sequence of stimuli interleaved with blanks. Prior work has posited that sustained attention can be enhanced via filling in the temporal blanks that could cause attentional lapses and disrupt the maintenance of attention (Huang et al., 2022, 2024). For instance, gradual-onset CPT tasks employ images that gradually transition from one stimulus to the next (Esterman et al., 2013), and visual timing studies also utilize continuously varying stimuli (Hove et al., 2010, 2013; Gan et al., 2015; Iversen et al., 2015; Huang et al., 2018). Using continuously varying instead of discrete stimuli acts as a form of interpolation between temporally separate events, though this approach introduces a confounding factor: continuously varying stimuli may carry more sources of information than sustained attention (i.e., increased motion-related information during stimulus variation versus increased attention). Subsequent studies resolved this confound by demonstrating a pure temporal filling-in effect without relying on continuously varying stimuli (e.g., by simply modifying the color of static sequence stimuli during blank periods to prevent attentional lapses caused by target offset) (Huang et al., 2022, 2024).

Collectively, temporal filling-in is proposed to improve sustained attention by mitigating attentional lapses, or in other words, reducing attentional fluctuations over time. This proposal remains unvalidated, however, given the difficulty of behaviorally tracking fluctuations in attentional states. To resolve this gap, the present study adopted a visual timing task that assesses the temporal filling-in effect (Fig. 1A) (Huang et al., 2022) and recorded visual steady-state evoked potentials (SSEPs) via electroencephalography (EEG) to track neural responses associated with attentional fluctuations. During rhythmic temporal processing, attentional states periodically rise and fall at the metronome beat frequency (Large and Jones, 1999) and can be tracked by a train of visual evoked potentials that are stable in amplitude and phase locked to the beat (i.e., SSEPs) (Norcia et al., 2015). SSEPs have been previously deployed to tag neural entrainment to auditory beats (Nozaradan et al., 2011), consistent with the dynamic attending theory (Large and Jones, 1999). We hypothesized that the sustained visual attention benefit of temporal filling-in would be associated with attenuated periodic attentional fluctuations, indexed by decreased amplitudes of visual SSEPs entrained to the visual beat (Fig. 1B).

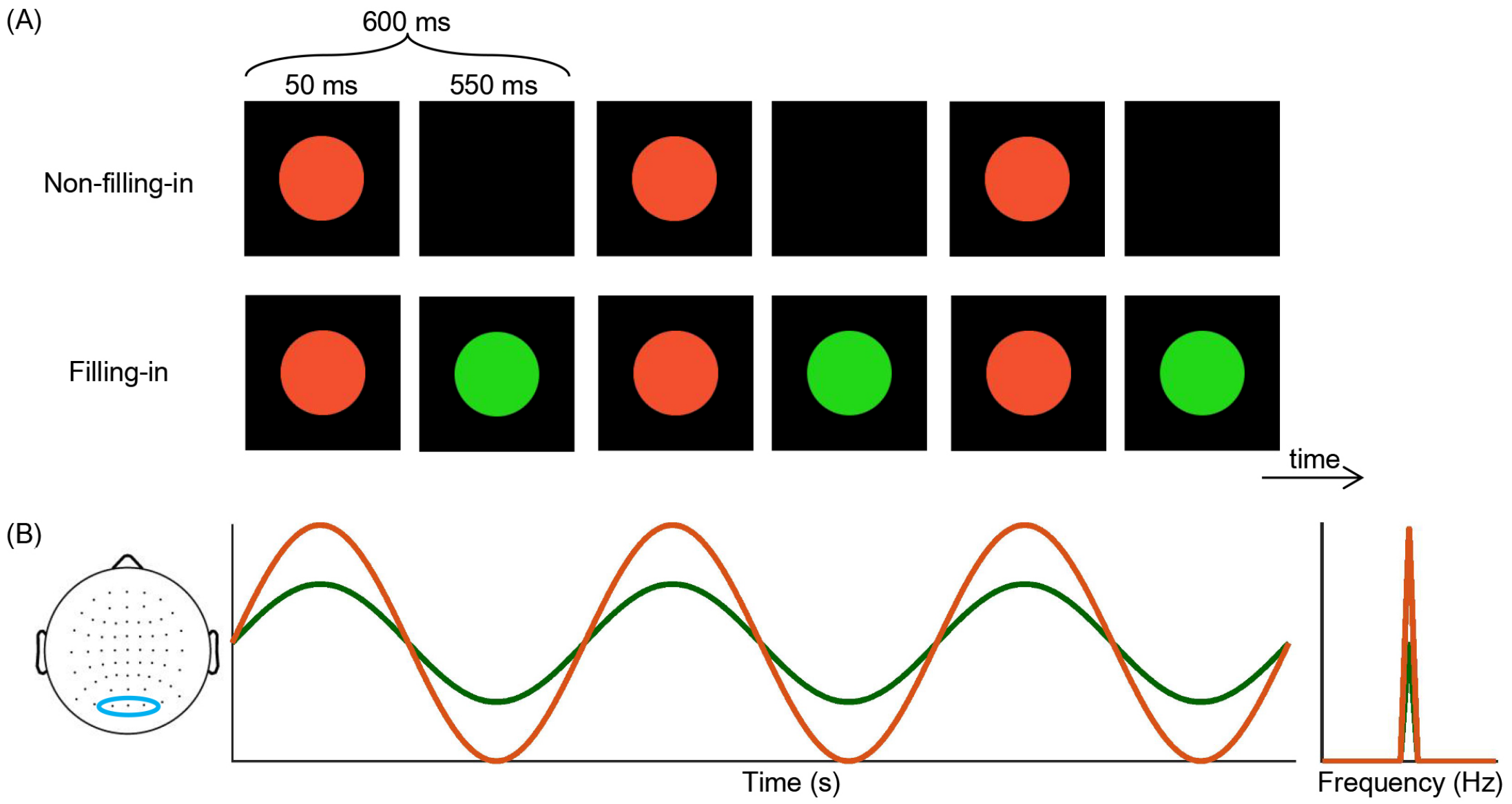

**Fig. 1. Illustration of the study design**. **A**. Experimental stimuli and procedure. Subjects were required to tap along with a visual metronome consisting of flashing red disks with a 600 ms inter-onset interval (IOI), i.e., beat interval (three cycles are shown). In the non-filling-in condition (top), the target red disk disappeared during the interstimulus blanks. In the filling-in condition (bottom), the temporal blanks were filled in by changing the disk's color to green. **B**. Diagram of the hypothesized reduction of attentional fluctuations under temporal filling-in. SSEPs recorded from occipital scalp areas (electrodes O1, Oz, O2, marked by the blue circle) indicate the periodic attentional fluctuations entrained to the metronome beat, visualized in the time and frequency domains (beat frequency: 1.66 Hz). Temporal filling-in reduces periodic attentional fluctuations, manifesting as decreased SSEP amplitudes (red: non-filling-in; green: filling-in).

## Methods

### Subjects

Twenty-five subjects (all right-handed; 4 males; 18-35 years, mean ± SD = 21.9 ± 3.2) at Sun Yat-Sen University participated in this study. All subjects had normal or corrected-to-normal vision.

### Data acquisition

#### Behavioral tasks

The task paradigm has been fully described in Huang et al. (2022); we summarize the experimental procedure briefly below. The subjects were instructed to tap in synchrony with an isochronous visual sequence (i.e., a visual metronome) by pressing a keyboard key with their right index finger. Two experimental conditions were implemented. In the non-filling-in condition, the sequence consisted of a brief 50 ms red flashing disk and a 550 ms blank interval. The inter-onset interval (IOI) was 600 ms and the corresponding beat frequency was 1.66 Hz. In the filling-in condition, temporal blanks were filled by changing the target disk's color to green. Each sequence had 40 target disk events and was repeated 10 times. Sequences were separated by blank rest intervals randomized between 3.3-4.9 s. The order of conditions was counterbalanced across subjects.

#### EEG data recording

EEG data were recorded during task performance using a 64-channel BioSemi ActiveTwo system (BioSemi, Amsterdam, Netherlands) with active electrodes positioned according to the extended 10–20 system, digitized at 2048 Hz. Horizontal and vertical electrooculography (HEOG and VEOG) were recorded to monitor eye movements.

## Data analyses

### Behavioral data

Behavioral data processing followed the routine described in Huang et al. (2022). Raw tapping data were processed using the circular analysis method suitable for variable periodic synchronization data (Fisher, 1993). Asynchrony was defined as the difference between the time of a tap and the time of the corresponding event onset, and was measured as the relative phase (RP) on a unit circle ($-\pi$ to $\pi$. Negative and positive values indicated taps preceding or following events, respectively). Successful synchronization for a sequence was assessed via the Rayleigh test of uniform distribution of the RPs. Synchronization stability was indexed by R, which was the length of the resultant (i.e., average of vectors) of the RPs and was calculated by abs(sum(exp(i*RP))/n) (n indicated the number of the RPs). R ranges from 0 (unstable tapping with uniformly distributed RPs) to 1 (perfectly stable tapping with a unimodal distribution of RPs). Correspondingly, mean asynchrony was indexed by the angle of the resultant of the RPs and was calculated by angle (sum(exp(i*RP))/n). Mean asynchrony analysis was restricted to successful sequences determined by the Rayleigh test. Because the stability is an indicator of the distribution of the RPs, all sequences (both successful and unsuccessful) were retained for stability analysis (Hove et al., 2013; Mu et al., 2018). Analyses focused on the stability, as it exhibits greater sensitivity to individual differences in synchronization performance relative to mean asynchrony. Taps corresponding to the first five events in a sequence were excluded from analyses because synchronization typically requires a few taps to stabilize (Gan et al., 2015; Iversen et al., 2015).

### EEG data

EEG data were analyzed offline using a custom processing routine involving MATLAB (for basic signal and statistical processing), EEGLAB (for EEG data preprocessing), Brainstorm (for EEG source reconstruction), and mfeeg (for basic EEG signal processing).

### EEG data preprocessing

Raw EEG data were downsampled to 500 Hz, 0.5 Hz high-pass filtered, and notch-filtered at 50/100/150/200/250 Hz. Bad electrodes were determined via EEGLAB's clean_rawdata procedure and interpolated with data from neighboring electrodes. The Infomax Independent Components Analysis (ICA) module in EEGLAB was used to decompose the EEG and remove artifact components. The data were re-referenced to the common average reference. Source reconstruction was conducted using the distributed models implemented in Brainstorm (Tadel et al., 2011). Source signal magnitude was quantified using the sLORETA normalized unit. A surface Brodmann atlas with 26 Brodmann areas (BAs) was adopted to parcellate the cortical surface (Fischl et al., 2008), and source time series of the brain regions were extracted by averaging vertex time series in each region. The EEG was then epoched for the sequence from the onset of the sixth stimulus to the end of the final stimulus (with 1 s padding on both sides to suppress potential boundary artifacts in later processing). The epochs were z-scored with the whole epoch length as the baseline, then averaged across sequences to obtain averaged sequence responses.

### SSEP analyses

#### Time domain

The averaged sequence responses were bandpass filtered at the beat frequency (1.667±0.5 Hz) to obtain the beat-specific sequence responses (Nozaradan et al., 2013; Łabęcki et al., 2024). To account for differences in global dispersions of beat-specific responses across conditions, we implemented a global sum-standard (sum-SD) normalization: each electrode’s beat-specific sequence response was divided by the summed standard deviation across all electrodes (the resulting collective standard deviations across electrodes summed to unity). Then a modeling analysis was conducted to obtain the amplitude and phase of the beat-specific sequence response.

Periodic attentional fluctuations were modeled as cosine waveforms spanning each beat cycle, with the amplitude and phase set as free fitting parameters. This theoretical sequence response was fitted to the empirical sequence response, yielding the modeled sequence response with optimal fitted amplitude and phase values.

#### Frequency domain

Frequency spectra were computed via Fast Fourier Transform (FFT) applied to the averaged sequence responses. Spectral amplitude at each frequency bin was divided by the average amplitude of eight neighboring frequency bins on either side (Gu et al., 2020). To account for differences in global beat-specific spectral amplitudes across conditions, a global sum normalization was implemented: each electrode’s beat-specific spectral amplitude was divided by the summed amplitude across all electrodes (the resulting collective beat-specific spectral amplitude across electrodes summed to unity).

### Statistical tests

All statistical tests used in this study are fully detailed in the main text and tables. In brief, Wilcoxon Signed-Rank tests were two-tailed; correlation analyses were performed using the Spearman correlation method. The significance threshold of $p<.05$ was adopted, with Bonferroni correction applied where appropriate.

## Results

### Behavior temporal filling-in effect

Task performance was measured primarily via synchronization stability. Synchronization was significantly more stable in the filling-in than in the non-filling-in condition (non-filling-in: mean = .55, 95% *CI* = [.47, .62]; filling-in: mean = .64, 95% *CI* = [.55, .73]; mean diff = .09, 95% *CI* = [.002, .19], $z = 2.06$, $p = .040$) (Fig. 2), replicating the temporal filling-in effect originally reported in Huang et al. (2022). In addition, mean asynchrony did not differ significantly between conditions (non-filling-in: mean = .03π, 95% *CI* = [−.16π, .21π]; filling-in: mean = −.17π, 95% *CI* = [−.26π, −.07π]; mean diff = −.19π, 95% *CI* = [−.39π, .01π], $z = 1.87$, $p = .062$).

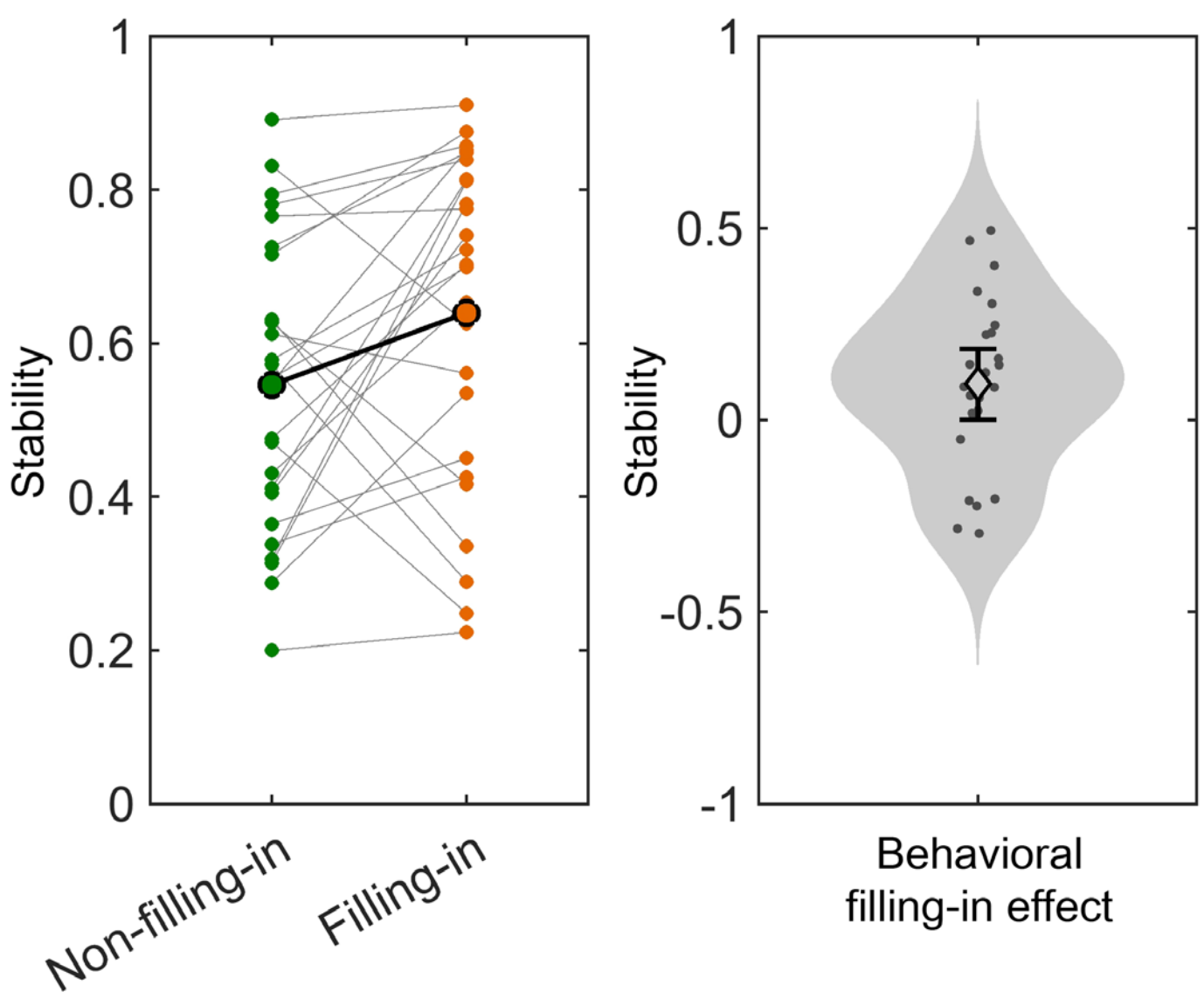

**Fig. 2. Illustration of behavioral results**. Left: 19 of 25 subjects tapped more stably in the filling-in than in the non-filling-in condition. Right: Group averaged filling-in minus non-filling-in stability difference data. Error bars denote ±95% CIs. Small dots indicate data from individual subjects.

## Attenuated visual SSEPs indexed reduced attentional fluctuations

### Analysis in the time domain

Responses in the occipital scalp area well captured neural entrainment to the visual beat, reflecting periodic fluctuations in visual attentional state (Fig. 3 A-D). The amplitudes of beat-specific sequence responses at the occipital electrodes O1, Oz, and O2 significantly decreased in the filling-in relative to the non-filling-in condition ($p < .001$) (Table 1). In addition, the phases of beat-specific sequence responses did not significantly differ between conditions ($p > .05$).

### Analysis in the frequency domain

Frequency-domain analyses yielded consistent results. Spectral amplitudes at the beat frequency at the occipital electrodes significantly decreased in the filling-in relative to the non-filling-in condition ($p < .01$) (Fig. 3 E-G).

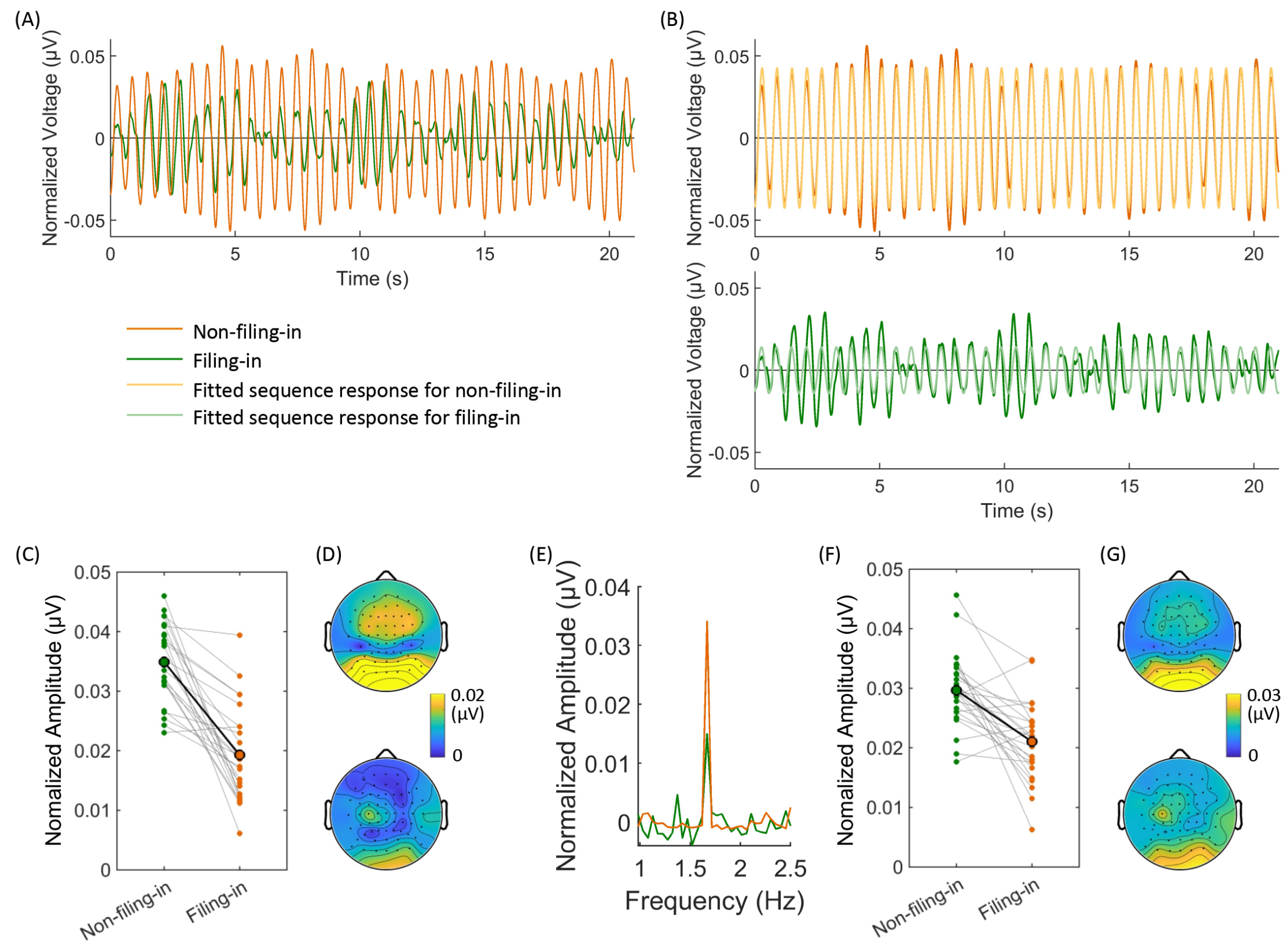

**Fig. 3. Illustration of visual SSEP results. A-D** present time-domain SSEP data. (**A**) illustrates beat-specific sequence responses (narrow-band filtered at beat frequency) for the non-filling-in (red) and filling-in (green) conditions at the occipital electrode Oz from one representative subject. (**B**) The amplitude and phase of the empirical sequence response were computed by fitting to the theoretical sequence response modeled as cosine waveforms spanning each beat cycle. (**C**) Individual time-domain beat-specific SSEP amplitudes. (**D**) further shows topographic maps for the time-domain beat-specific SSEP amplitudes averaged across subjects, with maximal amplitudes over the occipital region. **E-G** present frequency-domain data. (**E**) Frequency spectra at Oz from the representative subject. (**F**) Individual spectral amplitudes at beat frequency. (**G**) Topographic maps for the beat-specific spectral amplitudes averaged across subjects. Other conventions are as in Fig. 1 and 2.

| | Non-filling-in | | Filling-in | | Filling-in minus Non-filling-in | | | |
|---|---|---|---|---|---|---|---|---|
| Electrode | mean | 95% *CI* | mean | 95% *CI* | mean | 95% *CI* | *z* | $p_{corrected}$ |
| Time domain amplitude | | | | | | | | |
| O1 | .034 | [.012, .016] | .017 | [.008, .012] | -.016 | [-.020, -.013] | 4.37 | <.001 |
| Oz | .035 | [.012, .016] | .019 | [.008, .012] | -.016 | [-.020, -.012] | 4.35 | <.001 |
| O2 | .036 | [.012, .016] | .021 | [.008, .012] | -.015 | [-.019, .011] | 4.24 | <.001 |
| Time domain phase | | | | | | | | |
| O1 | .85π | [.79π, .90π] | .90π | [.74π, 1.06π] | .06π | [−.10π, .22π] | .50 | 1.856 |
| Oz | .85π | [.79π, .91π] | .92π | [.75π, 1.09π] | .07π | [−.09π, .23π] | .85 | 1.190 |
| O2 | .85π | [.79π, .92π] | .89π | [.74π, 1.04π] | .04π | [−.10π, .18π] | .20 | 2.520 |
| Frequency domain amplitude | | | | | | | | |
| O1 | .029 | [.027, .031] | .021 | [.018, .024] | -.008 | [-.012, -.004] | 3.43 | .002 |
| Oz | .030 | [.027, .032] | .021 | [.018, .024] | -.009 | [-.012, -.005] | 3.70 | .001 |
| O2 | .030 | [.028, .032] | .024 | [.021, .027] | -.006 | [-.009, -.003] | 3.24 | .004 |

**Table 1. Statistical values for analyses of visual SSEPs.** Conventions are as in Fig. 3.

**Motor SSEPs predicted behavioral performance**

In the present task, the processed visual information is utilized by the motor area to generate motor signals controlling synchronized tapping behavioral responses (Hove et al., 2013; Nozaradan et al., 2013). We further attempted to examine a potential link between motor responses and behavioral temporal filling-in effects, by computing Spearman correlations between the filling-in versus non-filling-in SSEP measurement differences at motor scalp electrodes and the behavioral stability differences. We targeted the left-central electrode C3 as well as its surrounding electrodes (FC3, C1, C5 and CP3), as the subjects tapped with their right hand (Nozaradan et al., 2013). For the time domain, the analyses showed a significant correlation for C3 beat-specific SSEP amplitudes ($p_{corrected}$=.007) (Fig. 4B) (Table 2). No significant correlation was found for SSEP phases. For frequency-domain beat-specific SSEP amplitudes, the analyses also revealed a significant correlation at C3 ($p_{corrected}$=.020) (Fig. 4C).

In addition, unlike visual SSEPs (attenuated under filling-in), motor SSEPs were elevated in the filling-in relative to the non-filling-in condition (Fig. 4D-G). For the time domain, beat-specific SSEP amplitudes significantly increased in the filling-in condition at FC3 ($p_{corrected}$=.047). Beat-specific SSEP phases did not differ significantly. Frequency-domain beat-specific SSEP amplitudes significantly increased in the filling-in condition at C3 ($p_{corrected}$=.007), C1 ($p_{corrected}$=.019), and CP3 ($p_{corrected}$=.002). The sensitivity differences between time-domain and frequency-domain analyses were likely related to their differing frequency resolution at the beat frequency: a 1 Hz bandwidth for narrowband time-domain filtering and a .049 Hz frequency bin width for frequency spectrum analyses.

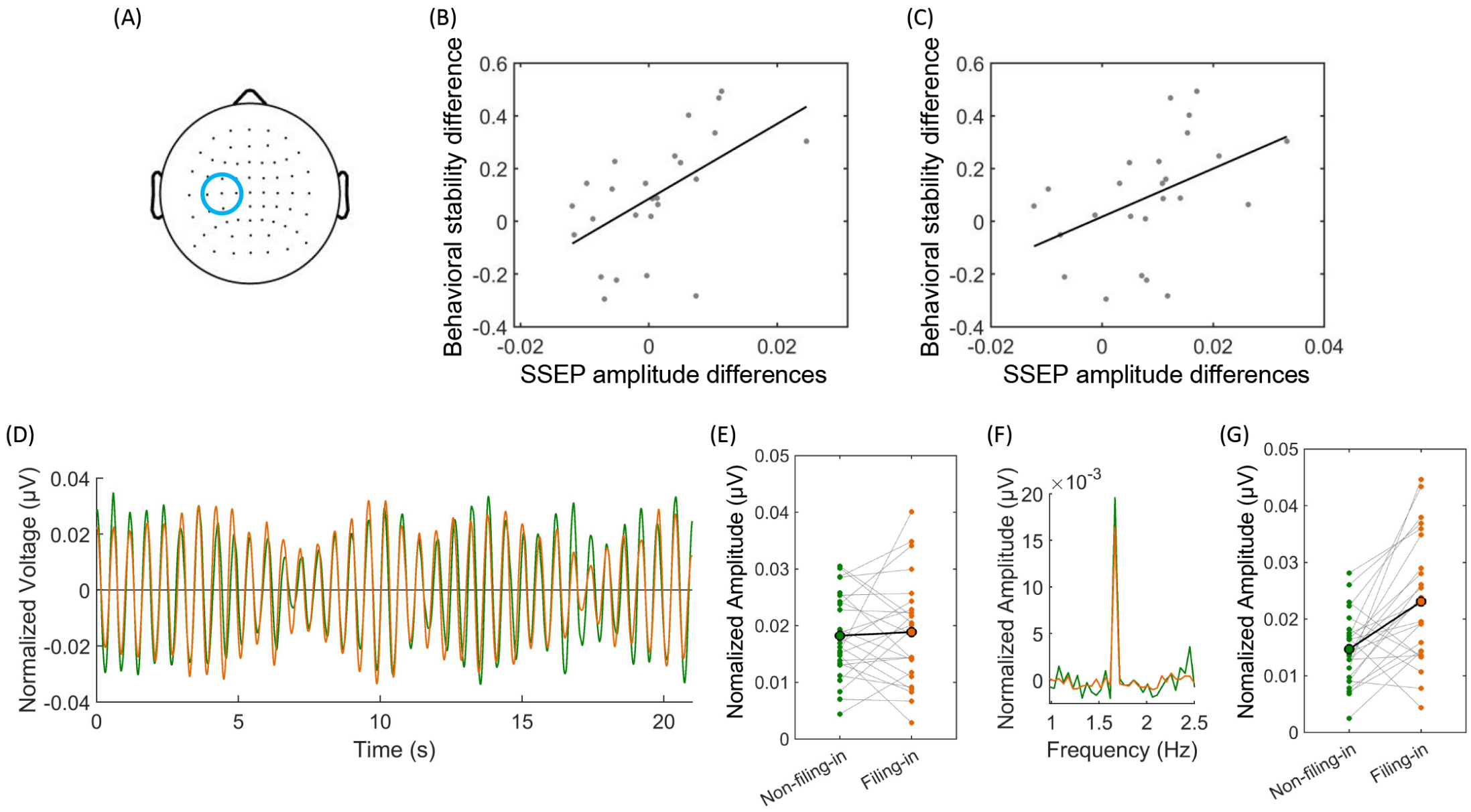

**Fig. 4. Illustration of motor SSEP results**. (**A**) Scalp montage highlighting the left motor electrodes (FC3, C1, C3, C5 and CP3; marked with the blue circle). (**B**) Scatter plot of behavioral temporal filling-in effects versus time-domain beat-specific SSEP amplitude differences at C3, with regression fit line. (**C**) Scatter plot for frequency-domain beat-specific SSEP amplitudes at C3. (**D**) Time-domain beat-specific sequence responses at C3 from a representative subject. (**E**) Individual time-domain beat-specific SSEP amplitudes. (**F**) Frequency spectra at C3 from the representative subject. (**G**) Individual spectral amplitudes at beat frequency. Other conventions are as in Fig. 3.

**Correlations of behavioral stability differences with SSEP measurement differences between conditions**

| | Time domain amplitude | | Time domain phase | | Frequency domain amplitude | |
|---|---|---|---|---|---|---|
| Electrode | *r* | $p_{corrected}$ | *r* | $p_{corrected}$ | *r* | $p_{corrected}$ |
| FC3 | .47 | .090 | -.07 | 3.779 | .42 | .181 |
| C1 | .50 | .059 | .17 | 2.032 | .51 | .054 |
| C3 | .62 | .007 | -.07 | 3.640 | .56 | .020 |
| C5 | .09 | 3.299 | .24 | 1.281 | -.05 | 4.048 |
| CP3 | .33 | .529 | -.35 | .438 | .42 | .187 |

**Comparisons of SSEP measurements between conditions**

| | Non-filling-in | | Filling-in | | Difference | | | |
|---|---|---|---|---|---|---|---|---|
| Electrode | mean | 95% *CI* | mean | 95% *CI* | mean | 95% *CI* | *z* | $p_{corrected}$ |
| Time domain amplitude | | | | | | | | |
| FC3 | .020 | [.012, .016] | .015 | [.008, .012] | -.005 | [-.009, -.001] | 2.60 | .047 |
| C1 | .018 | [.012, .016] | .016 | [.008, .012] | -.002 | [-.006, .002] | 1.01 | 1.565 |
| C3 | .018 | [.012, .016] | .019 | [.008, .012] | <.001 | [-.003, .004] | .09 | 4.625 |
| C5 | .013 | [.012, .016] | .013 | [.008, .012] | <.001 | [-.003, .003] | .07 | 4.732 |
| CP3 | .010 | [.012, .016] | .015 | [.008, .012] | .005 | [.001, .008] | 2.54 | .055 |
| Time domain phase | | | | | | | | |
| FC3 | −.13π | [−.22π, −.04π] | −.28π | [−.56π, −.01π] | −.16π | [−.42π, .11π] | .01 | 4.947 |
| C1 | −.15π | [−.26π, −.03π] | −.20π | [−.44π, .04π] | −.05π | [−.28π, .18π] | .39 | 3.482 |
| C3 | −.18π | [−.31π, −.05π] | −.18π | [−.38π, .02π] | .003π | [−.23π, .23π] | .42 | 3.383 |
| C5 | −.14π | [−.24π, −.05π] | −.21π | [−.36π, −.06π] | −.06π | [−.20π, .07π] | .17 | 4.306 |
| CP3 | −.07π | [−.25π, .11π] | −.19π | [−.37π, −.02π] | −.13π | [−.39π, .13π] | .71 | 2.379 |
| Frequency domain amplitude | | | | | | | | |
| FC3 | .016 | [.014,.018] | .018 | [.014, .021] | .001 | [-.003, .005] | .42 | 3.383 |
| C1 | .014 | [.011, .016] | .018 | [.015, .021] | .005 | [.001, .008] | 2.89 | .019 |
| C3 | .015 | [.012, .017] | .023 | [.019, .028] | .008 | [.004, .013] | 3.19 | .007 |
| C5 | .012 | [.010, .015] | .018 | [.013, .022] | .006 | [.001, .011] | 1.92 | .272 |
| CP3 | .009 | [.007, .011] | .017 | [.014, .021] | .008 | [.005, .012] | 3.51 | .002 |

**Table 2. Statistical values for analyses of motor SSEPs.** Conventions are as in Fig. 4.

### Analyses of source-reconstructed SSEPs

We further reconstructed scalp EEG data onto cortical source regions with a surface Brodmann atlas (Fischl et al., 2008) to examine whether the visual and motor SSEP results observed in scalp data could be accounted for by cortical source activity. We focused on primary and secondary visual and motor areas, including bilateral BA 17 (V1: primary visual cortex), bilateral BA 18 (V2: secondary visual cortex), left BA 4p (posterior primary motor cortex), left BA 4a (anterior primary motor cortex), and left BA 6 (premotor cortex) (Fig. 5; Tables 3 and 4).

For visual areas, time-domain beat-specific SSEP amplitudes significantly decreased under filling-in in bilateral BAs 17 and 18 ($p_{corrected}$<.001). Beat-specific SSEP phases did not differ significantly, although there was a marginal difference in left BA 18 ($p_{corrected}$=.053). Frequency-domain beat-specific SSEP amplitudes significantly decreased under filling-in in right BA 17 ($p_{corrected}$<.001), left BA 18 ($p_{corrected}$=.034), and right BA 18 ($p_{corrected}$<.001).

For motor areas, behavioral stability differences between conditions and time-domain beat-specific SSEP amplitude differences were marginally significantly correlated in left BA 4a ($p$=.067). Note that visual inspection of the correlation data revealed an outlier (Fig. 5K); excluding this outlier yielded significant correlations for both time-domain ($r$=.53, $p$=.008) and frequency-domain ($r$=.42, $p$=.042) beat-specific SSEP amplitudes. For comparisons of SSEP measurements between conditions, time-domain beat-specific SSEP amplitudes marginally significantly increased under filling-in in left BA 4p ($p$=.098); and SSEP phases differed significantly between conditions in left BA 4a ($p$=.002). Frequency-domain beat-specific SSEP amplitudes significantly increased under filling-in in left BA 4a ($p$=.042), left BA 4p ($p$<.001) and BA 6 ($p$<.001).

Together, consistent with the scalp data, these source-space results further revealed (1) decreased visual cortical SSEP amplitudes during filling-in within V1 and V2, and (2) increased motor SSEP amplitudes under filling-in and their correlations with behavioral performance differences within primary as well as secondary motor regions.

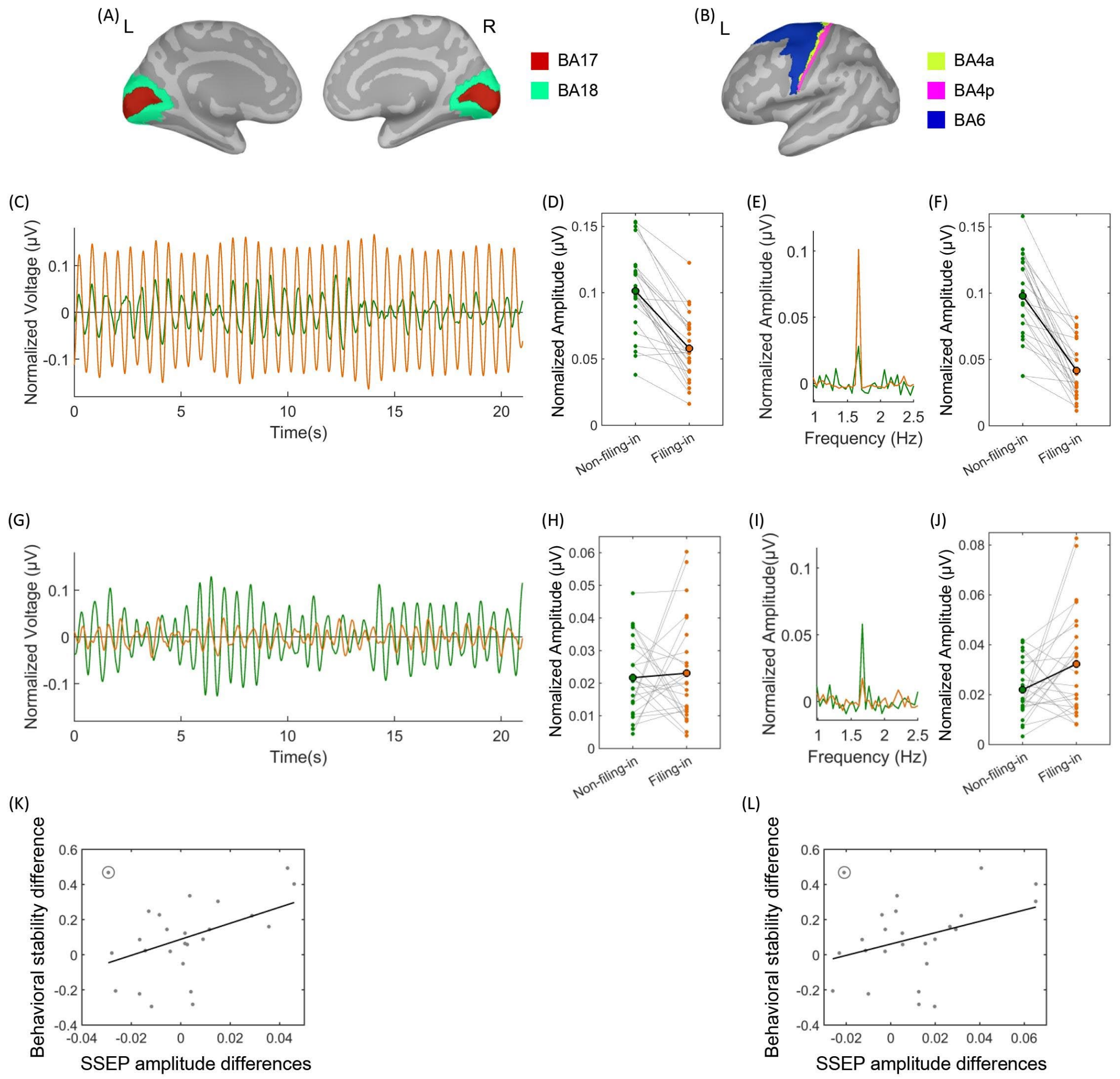

**Fig. 5. Illustration of source-reconstructed SSEP results.** (**A**) Bilateral visual areas BAs 17 and 18 on a surface brain template. (**B**) Left motor areas BAs 4a, 4p, and 6. **C**-**D** present data for the right BA17. (**C**) Time-domain beat-specific sequence responses from a representative subject. (**D**) Individual time-domain beat-specific SSEP amplitudes. (**E**) Frequency spectra from the representative subject. (**F**) Individual spectral amplitudes at beat frequency. **G-L** present data for the left BA 4a. (**G**) Time-domain beat-specific sequence responses from a representative subject. (**H**) Individual time-domain beat-specific SSEP amplitudes. (**I**) Frequency spectra from the representative subject. (**J**) Individual spectral amplitudes at beat frequency. (**K**) Scatter plot of behavioral temporal filling-in effects versus time-domain beat-specific SSEP amplitude differences. An outlier was marked by the circle. (**L**) Scatter plot for frequency-domain beat-specific SSEP amplitudes. Other conventions are as in Fig. 3 and 4.

| Region | Non-filling-in mean | Non-filling-in 95% *CI* | Filling-in mean | Filling-in 95% *CI* | Difference mean | Difference 95% *CI* | *z* | $p_{corrected}$ |
|---|---|---|---|---|---|---|---|---|
| Time domain amplitude | | | | | | | | |
| BA17 L | .072 | [.019, .029] | .042 | [.018, .030] | -.030 | [-.043, -.017] | 3.73 | <.001 |
| BA17 R | .098 | [.019, .029] | .042 | [.018, .030] | -.056 | [-.069, -.044] | 4.37 | <.001 |
| BA18 L | .076 | [.019, .029] | .036 | [.018, .030] | -.041 | [-.054, -.027] | 4.08 | <.001 |
| BA18 R | .097 | [.019, .029] | .043 | [.018, .030] | -.054 | [-.068, -.040] | 4.24 | <.001 |
| Time domain phase | | | | | | | | |
| BA17 L | .96π | [.74π, 1.17π] | .96π | [.74π, 1.18π] | .001π | [−.37π, .37π] | .31 | 1.514 |
| BA17 R | 1.13π | [.96π, 1.30π] | 1.04π | [.86π, 1.24π] | −.08π | [−.33π, .16π] | .63 | 1.054 |
| BA18 L | 1.28π | [1.00π, 1.57π] | .95π | [.71π, 1.20π] | −.32π | [−.64π, −.01π] | 2.22 | .053 |
| BA18 R | .99π | [.80π, 1.19π] | .93π | [.72π, 1.13π] | −.06π | [−.34π, .21π] | .63 | 1.054 |
| Frequency domain amplitude | | | | | | | | |
| BA17 L | .075 | [ .062, .090] | .060 | [.051, .070] | -.016 | [-.031, -.001] | 1.63 | .207 |
| BA17 R | .101 | [.089, .114] | .058 | [.048, .069] | -.043 | [-.053, -.033] | 4.35 | <.001 |
| BA18 L | .078 | [.062, .093] | .054 | [.042, .066] | -.024 | [-.042, -.005] | 2.38 | .034 |
| BA18 R | .093 | [.081, .105] | .061 | [.045, .078] | -.032 | [.049, -.015] | 3.43 | <.001 |

**Table 3. Statistical values for analyses of source-reconstructed visual SSEPs.** Conventions are as in Table 1 and Fig. 5.

**Correlations of behavioral stability differences with SSEP measurement differences between conditions**

| Region | Time domain amplitude *r* | Time domain amplitude *p* | Time domain phase *r* | Time domain phase *p* | Frequency domain amplitude *r* | Frequency domain amplitude *p* |
|---|---|---|---|---|---|---|
| BA4a L | .37 | .067 | -.02 | .937 | .33 | .102 |
| BA4p L | .19 | .361 | .06 | .759 | -.01 | .978 |
| BA6 L | -.02 | .917 | .17 | .421 | -.20 | .329 |

**Comparisons of SSEP measurements between conditions**

| Region | Non-filling-in mean | Non-filling-in 95% *CI* | Filling-in mean | Filling-in 95% *CI* | Difference mean | Difference 95% *CI* | *z* | *p* |
|---|---|---|---|---|---|---|---|---|
| Time domain amplitude | | | | | | | | |
| BA4a L | .022 | [.019, .029] | .023 | [.018, .030] | .001 | [-.007, .010] | .04 | .968 |
| BA4p L | .027 | [.019, .029] | .035 | [.018, .030] | .007 | [-.001, .015] | 1.65 | .098 |
| BA6 L | .022 | [.019, .029] | .026 | [.018, .030] | .004 | [-.002, .010] | 1.33 | .183 |
| Time domain phase | | | | | | | | |
| BA4a L | −.17π | [−.45π, .12π] | −.59π | [−.81π, −.36π] | −.42π | [−.67π, −.18π] | 3.08 | .002 |
| BA4p L | −.11π | [−.40π, .18π] | −.39π | [−.60π, −.19π] | −.28π | [−.66π, .09π] | 1.39 | .166 |
| BA6 L | −.13π | [−.37π, .11π] | −.37π | [−.59π, −.14π] | −.23π | [−.56π, .10π] | 1.28 | .201 |
| Frequency domain amplitude | | | | | | | | |
| BA4a L | .021 | [.017, .026] | .032 | [.023, .041] | .010 | [.001, .020] | 2.03 | .042 |
| BA4p L | .026 | [.019, .033] | .048 | [.037, .058] | .022 | [.012, .031] | 3.89 | <.001 |
| BA6 L | .022 | [.018, .027] | .040 | [.029, .051] | .018 | [.008, .028] | 3.51 | <.001 |

**Table 5. Statistical values for analyses of source-reconstructed motor SSEPs.**

Conventions are as in Table 2 and Fig. 5.

## Discussion

To test the hypothesis that temporal filling-in enhances sustained attention via reducing attentional fluctuations, the current study recorded visual SSEPs during a rhythmic visual timing task assessing the temporal filling-in effect. The results showed decreased visual SSEPs in the filling-in than in the non-filling-in condition, indexing the reduced fluctuations of visual attention. The results also showed increased motor SSEPs that correlated with behavioral benefits. Together, these data suggest that temporal filling-in reduces periodic rises and falls in visual attentional state, which may facilitate motor responses controlling behavioral responses.

It is worth noting that the color alternation of target stimuli in the filling-in condition may appear more salient, which would predict increased rather than decreased visual SSEPs (Tsuchiya and Koch, 2005). Our experimental design would thus make two competing predictions: the saliency account predicted elevated visual SSEPs, whereas the temporal filling-in account predicted attenuated visual SSEPs. The empirical data supported the filling-in account.

Attentional states wax and wane moment to moment, and it is challenging to directly measure the attentional fluctuations. To circumvent this difficulty, we employed a rhythmic visual timing task in which visual attention fluctuates periodically (Large and Jones, 1999) and used visual SSEPs to track visual responses entrained to the periodicity (Norcia et al., 2015). In this regard, two considerations should be taken into account when interpreting the findings. First, beyond its utility for probing periodic attentional dynamics, this visual timing paradigm and the associated temporal filling-in effect reflect the characteristics of sustained visual attention in general (extended theoretical discussion in Huang et al., 2024). Second, while the recorded visual SSEP reflected the effect of dynamic visual attention, it was not a direct measure of the source attention state itself; how to isolate the rhythmic attentional control from its downstream effect remains an open question. Relatedly, the motor SSEP indexed anticipatory motor control. Consistent with prior auditory timing work (Nozaradan et al., 2013), motor SSEPs exhibited positive phases (i.e., ahead of stimulus onsets). Though the decreased visual SSEPs and increased motor SSEPs indicate that reduced visual attentional fluctuations facilitate motor control processing, we emphasize that the current data did not resolve the dynamic interactions between the sensory and motor processes.

More broadly, the beneficial temporal filling-in effect on sustained attention via reduced attentional fluctuations aligns conceptually with a counterintuitive finding from Olivers and Nieuwenhuis (2005). In their attentional blink task detecting two temporally adjacent numbers embedded in letter streams, distracting from the task by task-irrelevant activity (e.g., free association) paradoxically improved rather than impaired task performance. Given how simple it is to detect a number, the pronounced deficit in identifying the second number would seem counterintuitive. The human cognitive system may have developed to support hyper-focused attention under high-stakes or threatening contexts - an adaptive strategy for survival-critical circumstances. Such hyperfocus, however, becomes inappropriate for cases such as simply detecting a number: excessive prioritization of the first target suppresses processing of the second, leading to the attentional blink deficit. From this viewpoint, the attentional blink paradigm itself generates the counterintuitive behavioral pattern, rather than the distracting mental activity: the distraction alleviates the deficit by mitigating unnecessary hyperfocus (Olivers and Nieuwenhuis, 2005). Analogously, processing rhythmic temporal information (e.g., watching dance) is a ubiquitous daily cognitive activity in which extreme attentional fluctuations (focus on discrete periodic time points) may represent a suboptimal strategy. Reducing these large attentional fluctuations via temporal filling-in therefore yields behavioral performance benefits.

This work has several limitations. We recorded SSEPs via scalp EEG, which is constrained by relatively low spatial resolution, although source-reconstructed data showed results consistent with the scalp data. The regions of interest (ROIs) for source analyses were predefined based on a brain atlas. While this atlas parcellation provides fine delineations, such as the separation between BAs 4p and 4a, an independent functional localizer targeted to the index finger used for task responses could improve the sensitivity of this investigation. Moreover, future work may leverage intracranial EEG (iEEG) to acquire cortical SSEPs with simultaneously high temporal and high spatial resolution (Mercier et al., 2022).

In conclusion, maintaining sustained attention is inherently limited by attentional fluctuations. Our results demonstrate that the sustained visual attention benefits conferred by temporal filling-in were associated with reduced attentional fluctuations as indexed by attenuated visual SSEP amplitudes. The findings reveal a fundamental mechanism for enhancing temporal attention via mitigating hyper-focused attentional states over time.

**Acknowledgments**

This work was supported by National Natural Science Foundation of China [31971033]. The study was performed via the research platform "11buddy & I" (https://github.com/rwfwuwx/11buddy-and-I).

**Ethical approval**

The research protocols and procedures in this study were approved by the Institutional Review Board of Psychology Department of Sun Yat-Sen University, and were in accordance with the 1964 Helsinki declaration and its later amendments or comparable ethical standards. All participants provided written informed consent before their participation.

**Conflict of interest**

The authors declare no conflict of interest.

**Data availability statement**

The data generated during and/or analyzed during the current study are available from the corresponding author on reasonable request.

**Author contributions**

Yingyu Huang: Methodology, Software, Formal analysis, Writing - Original Draft, Writing - Review & Editing. Liying Zhan: Methodology, Software. Xiang Wu: Conceptualization, Methodology, Software, Formal analysis, Writing - Original Draft, Writing - Review & Editing, Supervision, Funding acquisition.

**Declaration of generative AI and AI-assisted technologies in the manuscript preparation process**

During the preparation of this work, the authors used Doubao (ByteDance Inc.) for language polishing and error correction of the manuscript. Following use of the tool, the authors reviewed and edited the content as necessary, and accept full responsibility for the content of the published article.